# Optical Doppler shift measurement using a rotating mirror

Luis Bernal

Physics Department, Faculty of Sciences, University of Mar del Plata, Argentina

Luis Bilbao

Plasma Physics Institute (CONICET) and Physics Department, FCEN, University of

Buenos Aires, Argentina

Optical Doppler shift demonstration experiments are not a simple task since the light source cannot usually be moved in a sufficiently smooth and uniform manner to keep the level of noise well below of that of the signal. For that reason most demonstration experiments are usually performed with sound or with microwaves.

Previous works have been reported using a moving mirror in order to produce a moving light source, but small perturbation of its trajectory, as small as the optical wavelength, can produce a large noise.

Using a rotating mirror, in which one beam is reflected from the advancing side and the other beam is reflected from the receding part of a rotating mirror, can overcome many of the noise generating effects.

In the present work we report the construction and operation of a demonstration apparatus for measuring optical Doppler shift based on a rotating mirror.

01.50.My, 07.07.-a

# Introduction

Experiments on mechanical and electromagnetic wave propagation are usually conducted in elementary, introductory physics courses. Many teaching kits are readily available on the market covering most of the key experiments. Notwithstanding this fact neither those kits nor simple teaching experiments on optical Doppler shift are easily found.

Although the Doppler shift is a well-understood effect that is usually introduced to undergraduate physics students in elementary courses, its demonstration using light is not a simple matter. Doppler shift demonstration experiments for undergraduates are usually performed with sound or with microwaves.

Using light is not a simple task since the moving mirror from which light is reflected cannot usually be moved in a sufficiently smooth and uniform manner to keep the level of noise below that of the signal. Small perturbations as small as the optical wavelength can produce a non-negligible noise.

Previous undergraduate demonstrations of the optical Doppler shift have involved the use of Michelson interferometers with arms of approximately equal path lengths using air-track glider[1] or a servo-mounted mirror on an isolated optics table[2]; light reflection from Scotchlite tape on a rotating turntable with a spectrum analyzer[3]; and direct frequency modulation produced by a HO scale model train engine[4]. Velocities up to 0.3 m/s were used, thus Doppler beat frequencies were below 1 MHz.

One of the proposed goals of the present work was to reach a room-size beat wavelength that corresponds to a beat frequency of about 30 MHz or more. Therefore, velocities of the order of 10 m/s are required. Using linear translation techniques at 10 m/s or higher velocity in a standard classroom or laboratory is almost impossible. Therefore a rotational technique was devised in which one beam is reflected from the advancing side and the other beam is reflected from the receding part of a rotating mirror. An apparatus was built which students can use to measure the Doppler shift as a function of the rotating frequency of a rotating mirror.

**Doppler shift**

Light frequencies are around 500 THz, well beyond the frequency resolution of light detectors. Optical photodetectors are square law devices. Therefore, the measured intensity fluctuation is the temporal mean value of the optical perturbation over a time span related to the detector time response, always much larger than the period of the light wave. On the other hand, the mixing or beat frequency corresponds to a much lower frequency and can be detected using an adequate fast photodetector.

Two effective sources with different velocities can be created using a rotating mirror. The beat frequency between the two sources is related to the first order Doppler shift of the beam reflected by the rotating mirror.

A plane-polarized light wave reflected from a uniformly moving mirror will suffer a frequency shift.[5] Let $v_n$ be the velocity normal to its plane (positive towards the source) of a mirror in uniform motion, and a plane wavefront which makes the angle of incidence $\alpha$ with the mirror, then the frequency of the reflected beam, to first order in $v_n/c$ ($c$ is the velocity of light), will be[5] (a simple geometrical explanation can be found, for example, in Ref. 6)

$$\omega' = \omega\left(1 + 2\frac{v_n}{c}\cos\alpha\right) \qquad (1)$$

Actually, this relation follows simply from the consideration that the image of the source moves with the velocity $2v_n$ in the direction of the normal to the mirror, and, consequently, the component of this velocity in the direction of the reflected ray is $2v_n \cos\alpha$.

For a non-uniform movement of the mirror the same formula (1) is valid using the instantaneous normal velocity of the reflecting surface.[7]

Note that for the geometry of our experiment (see Fig. 1) the normal velocity $v_n$ is proportional to the distance $R$ from the midpoint of the reflecting surface, according to

$$v_n = v \frac{R}{L} \qquad (2)$$

where $v$ is the local velocity of the mirror and $L$ the distance from the axis of rotation. Since $v = \Omega L$ ($\Omega$ is the rotating angular frequency of the mirror), the frequency of a reflected beam at a distance $R$ from the midpoint of the reflecting surface will be (see Fig. 1)

$$\omega' = \omega\left(1 \pm 2\frac{R\Omega}{c}\cos\alpha\right) \qquad (3)$$

where the sign will be positive in the part where the mirror advances to the source and negative where it recedes from the source.

Taking two beams reflecting from opposite points from the midpoint of the reflecting surface, then

$$\omega_1 = \omega\left(1 + 2\frac{R\Omega}{c}\cos\alpha\right)$$
$$\omega_2 = \omega\left(1 - 2\frac{R\Omega}{c}\cos\alpha\right) \qquad (4)$$

The corresponding beat angular frequency of the beams is

$$\omega_b = \omega_1 - \omega_2 = \frac{4\omega R\Omega \cos\alpha}{c} \qquad (5)$$

Since the separation $d$ between the two beams centered at the midpoint of the reflecting surface is (see Fig. 1)

$$d = 2R\cos\alpha \qquad (6)$$

then, the frequency $\nu_b = \omega_b/2\pi$ can be written as

$$\nu_b = \frac{4\pi dF}{\lambda} \qquad (7)$$

where $F$ is the frequency of rotation of the rotating mirror.

Note that the beat frequency (7) depends on beam separation rather than the radial position where the beams hit the mirror. The same frequency will be obtained if, for example, both beams are at equal, opposite distance $R$ from the midpoint of the reflecting surface or one beam hits the mirror at its midpoint and the other at distance $2R$. That is, under this particular geometry, only beam separation counts.

Further, according to (7), the beat frequency does not depend on the angle of incidence either. If the angle of incidence of the parallel beam is larger, then the beam separation at the mirror surface will be larger by the same factor as the reduction of the component of the normal velocity in the direction of the reflected ray. Thus the angle of incidence cancels out.

Therefore the measurement is almost insensitive to perturbations to the mirror movement. Any standard grade rotating tool will be sufficient to obtain a good measurement of the optical Doppler shift. This is a clear advantage over translating movements were beams are not self compensated.

Note that the above is valid to the first order in $v_n/c$. At higher speed (not applicable to our experiment) second order effects need to be considered[5] including the modification of the angle of reflection (some examples of this relativistic effect can be seen in Ref. 8).

Since both, source and detector are at rest in the laboratory frame of reference, another simple, alternate way for obtaining (7) is to calculate the different path from the source to the detector for the two arms of the interferometer.[9] Using a computer algebra software package it is possible to obtain a more general derivation than this particular case. The interpretation as a path difference is usually easier to explain to undergraduate students. Also, using this approach it is easier to obtain the dependence of the intensity with time. The rotating beam is sweeping the photomultiplier slit, thus the intensity varies with time in a way similar to that from a lighthouse. Assuming a given beam diameter, it is possible to calculate a realistic time variation of the intensity, simply as the geometrical overlap of the sweeping circular beam and the slit.

**Experimental setup**

A block diagram of the apparatus is shown in Fig. 2. The light source is a low-power green He-Ne laser. A beam splitter separates the beam, part is reflected from the advancing side of a rotating mirror; part is reflected from the receding part of a rotating mirror. They are recombined by a beam splitter and measured by a photo-multiplier tube (PMT). Beat between the two beams produces fluctuations in the light intensity at the photo-cathode. The photocurrent is proportional to the light intensity and contains fluctuations due to the Doppler shift of the reflected beam from the rotating mirror.

The laser was a Melles Griot model 05-LGR-025. The laser beam was used with no lenses. The unmodified laser beam is easy to position and to control directionally. Although not needed for demonstration purposes, an optional spatial filter was designed to be placed at the laser output.

The rotating mirror was handcrafted by one of the authors on a cube (20-mm side) of Cobalt-Steel. Only one of the four faces was mirrored (front surface) producing one pulse per rotation. A low cost ($20), high-speed (up to 30,000 rpm) rotary tool was used to drive the rotating mirror. A special support was constructed to hold the rotatory tool. Beam-splitters and mirror were from standard optical Melles-Griot kits. Special positioners were built in order to place mirror and beamsplitter close enough to produce a 17 mm beam separation or recombination.

For detecting the beat frequency a Hamamatsu R928 photomultiplier powered by a 2KV, 2mA power supply, was used. The electrical signal was recorded in a 100 MHz, 500 MSa/s Tektronik TDS 320 digitizer or in a 500 MHz, 1 GSa/s Hewlett Packard Infinium (that has a built in Fast Fourier Transform). Photomultiplier cage was made of aluminum.

A photodiode connected to a low-frequency meter was used for measuring the rotating frequency.

All the components were mounted on a 0.3 m by 0.5 m iron base.

No special skills are needed for setting up the apparatus. The alignment is as difficult as that of a Mach-Zehnder interferometer. With the mirror at rest the alignment is performed as follows. Using the first mirror-beamsplitter set and removing the rotatory tool, a parallel beam is produced. Then the rotatory tool is set in place and the mirror is manually rotated until one of the reflected beams hits the photomultiplier slit. Then, by means of the second mirror-beamsplitter set the two reflected beams are recombined at the photomultiplier slit. The interference pattern between the two Airy disks can be easily produced on the slit. A final, minor adjustment can be performed while the mirror is rotating in order to get the best beating pattern.

At a given rotational speed the duration of the beating signal can be increased or decreased by locating the detector at different distance from the rotating mirror. In the present work we have obtained signal widths decreasing from 20 to 1 µs, by moving the PMT up to 6 m from the mirror.

Since the PMT is a very sensitive to light some care should be taken in order to prevent room light from reaching the photomultiplier. Depending on the configuration, some screening may be needed.

**Results**

In our experiment $\lambda = 543.5\ nm$ (green HeNe laser), and $d = 17\ mm$. The frequency of the mirror was varied from 20 to 200 Hz.

In order to illustrate a sample output in Fig. 3 we show the temporal variation of intensity for the case $F = 81.6\ Hz$. In this example the PMT was located at 4 m from the rotating mirror in order to give an idea of the full pulse including its Doppler modulation inside. At shorter distance the pulse width increases and since the Doppler beat frequency is the same it is very difficult to see both the pulse and its details in a fixed scale. The measured frequency $\nu = 32.1\ MHz$ is in good agreement with the expected value according to (7).

The measurements proceeded as follows. Using a variable transformer the speed of the rotating mirror was varied from 20 to 200 Hz. At approximately 10 Hz interval the photomultiplier signal was saved to the computer together with the photodiode measure of the mirror frequency (depending on the digitizer up to 32,768 data points were recorded). A Fast Fourier Transform (FFT) was applied to each signal and the peak frequency was recorded. In Fig. 4 the FFT corresponding to Fig. 3 is plotted.

Some digitizer have a built in FFT function, therefore a direct read can be obtained. Using 32,768 data points a typical FWHM of the FFT was smaller than 1% of the beat frequency, giving enough precision for demonstration purposes.

A plot of the peak FFT frequency as a function of the rotating frequency was constructed. In Fig. 5 we plot the measured beat frequency as a function of the rotating mirror frequency. Also the theoretical values are plotted with dotted line. The agreement is excellent. All experimental values agree to the theoretical values within the error limits.

## Discussion

The beat frequency difference between the Doppler shifted light from a rotating mirror was observed for different rotational frequencies. Using low precision rotatory tool very good results can be obtained.

All components that were used are commonly available items. The techniques described in this paper provide a quantitative demonstration of the Doppler effect of light and overcome the usual problems caused by motional instabilities associated with the moving mirror, since the frequency shift depends on beam separation rather than radial position or axial displacement.

Velocities of the moving mirror were up to 10 m/s, that is more than one order of magnitude larger than in previous work. This allows an optical beat frequency up to 80 MHz in a 1 to 20 µs pulse length. This pulse can be used to perform other demonstration measurements including the speed of light without needing special extra equipment since wavelength is expected to be from 4 to 15 meters.

Another advantage is that the experiment is not limited to a restricted temporal window as in methods that employ pure translation of the source (the mirror must start and stop at some point). One can continuously acquire data using a rotating mirror, which is especially useful for demonstration.

An interesting feature of our approach is that neither sophisticated optical nor electronics is needed. The main items required are a low-powered laser, mirrors, optical and detector. These items are available in most universities and colleges. As a tradeoff a relatively fast digitizer having 100 MHz bandwidth and 500 MSa/s plus high-speed photodetectors (in this case a PMT) are needed. According to present day technology these items are not highly demanding.

Of course the present method can in principle be used at lower rotational speed. We have not investigated this low speed limit since our low cost rotatory tool was unable to smoothly rotate below 10 Hz. As suggested by one of the reviewer, note that a rotational angular speed of earth rate (15°/hour) corresponds to a Doppler beat of about 8 Hz. This can be a very interesting way to show students how to detect very low rotational speeds.

## Acknowledgments

Authors wish to thank Professor F. Minotti for his suggestions. Work partially supported by grants UBA-X206, and CONICET PIP 5291/05.

## Figure captions

**Figure 1.** Reflection of two beams separated by a distance *d* centered at the midpoint of the reflecting surface. The normal velocity at any point of the mirrored surface is proportional to the distance to its midpoint.

**Figure 2.** Experimental arrangement. Arrowheads on light paths indicate direction of observed light. Arrows by rotating mirror represent rotation.

**Figure 3.** Sample oscilloscope output of the Doppler beat frequency produced by a rotating mirror at 81.6 Hz measured at 4 m from the mirror: a) extended plot, showing the time variation of the light intensity, at 250 ns/div, 200 MSa/s , b) expanded plot showing the beat pattern at 25 ns/div, 500 MSa/s.

**Figure 4.** 1,024 data points Fast Fourier Transform of the signal of Fig. 3a. (a) The full spectrum, (b) an enlargement around the main peak showing the frequency of the peak and its FWMH (depending on the digitizer up to 32,768 data points were recorded, thus reducing the FWMH).

**Figure 5.** Doppler beat frequency as a function of the rotational frequency of the mirror. Full line, theoretical values; squares, experimental values (error bars are within the square limits).

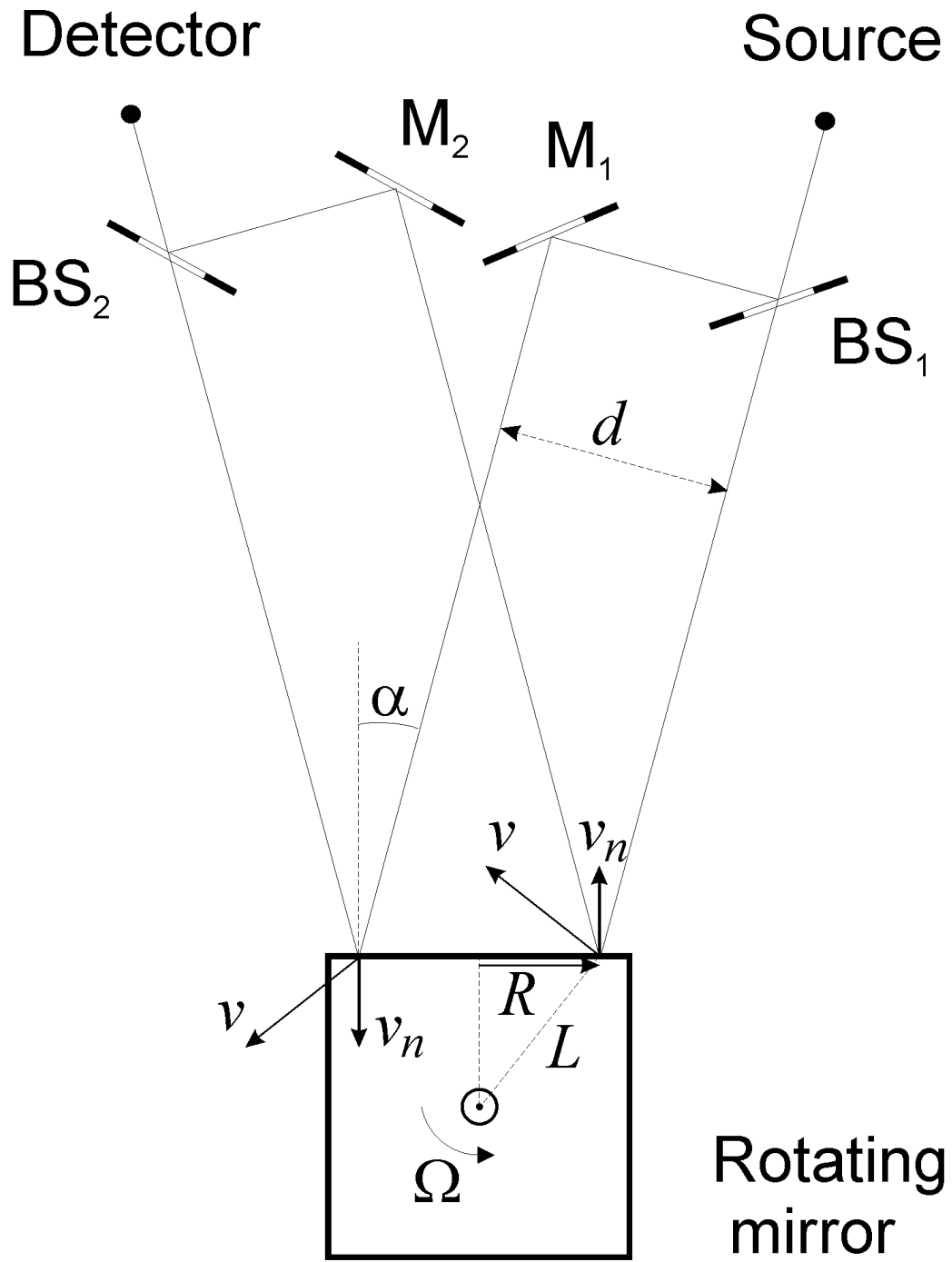

BernalFig1

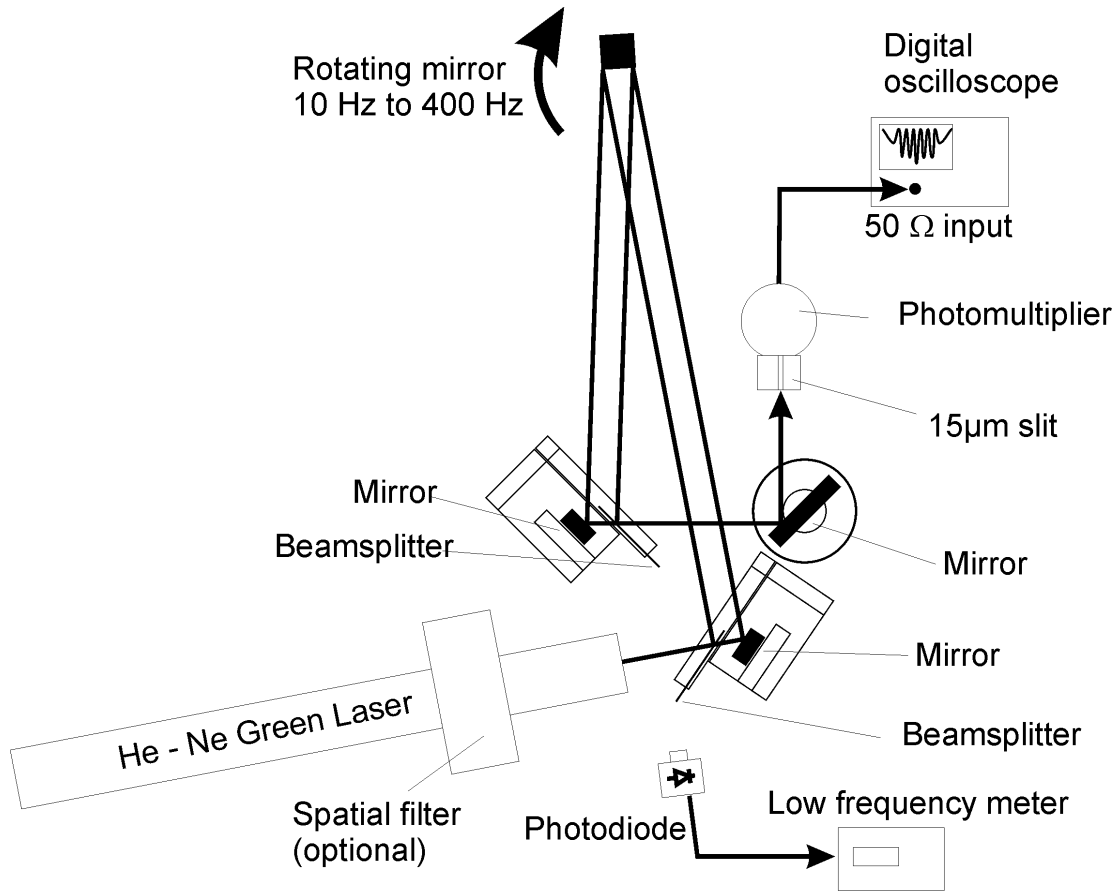

BernalFig2

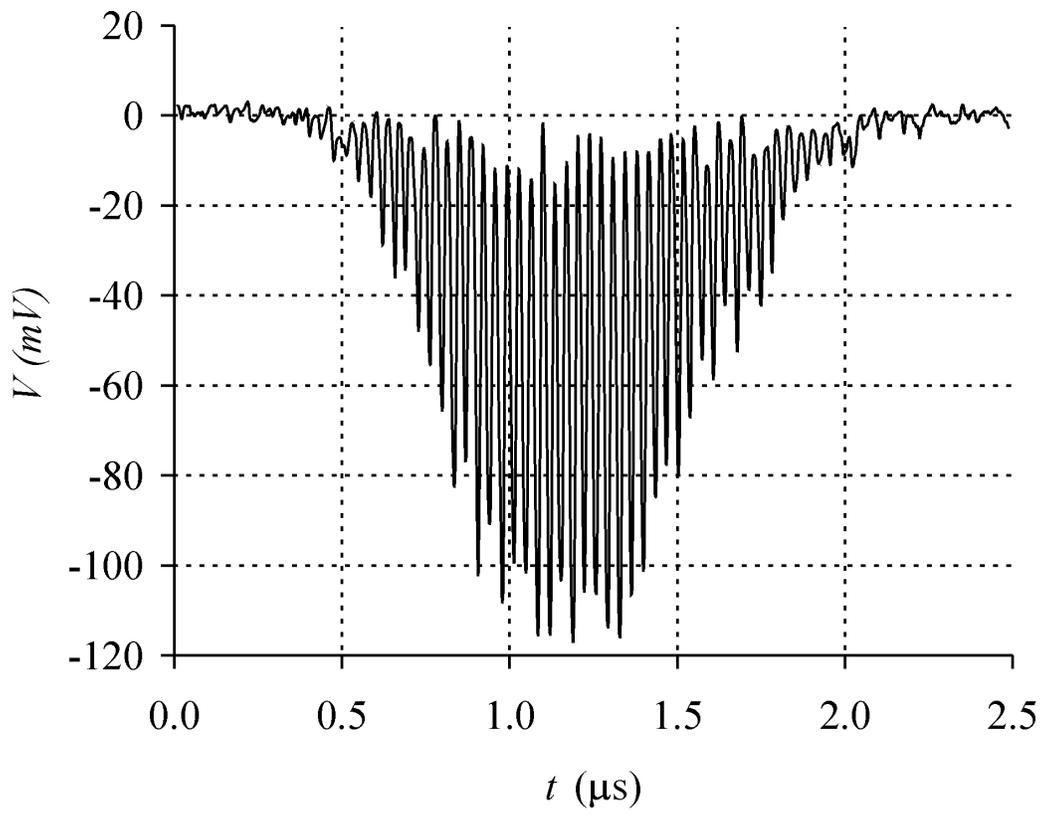

BernalFig3a

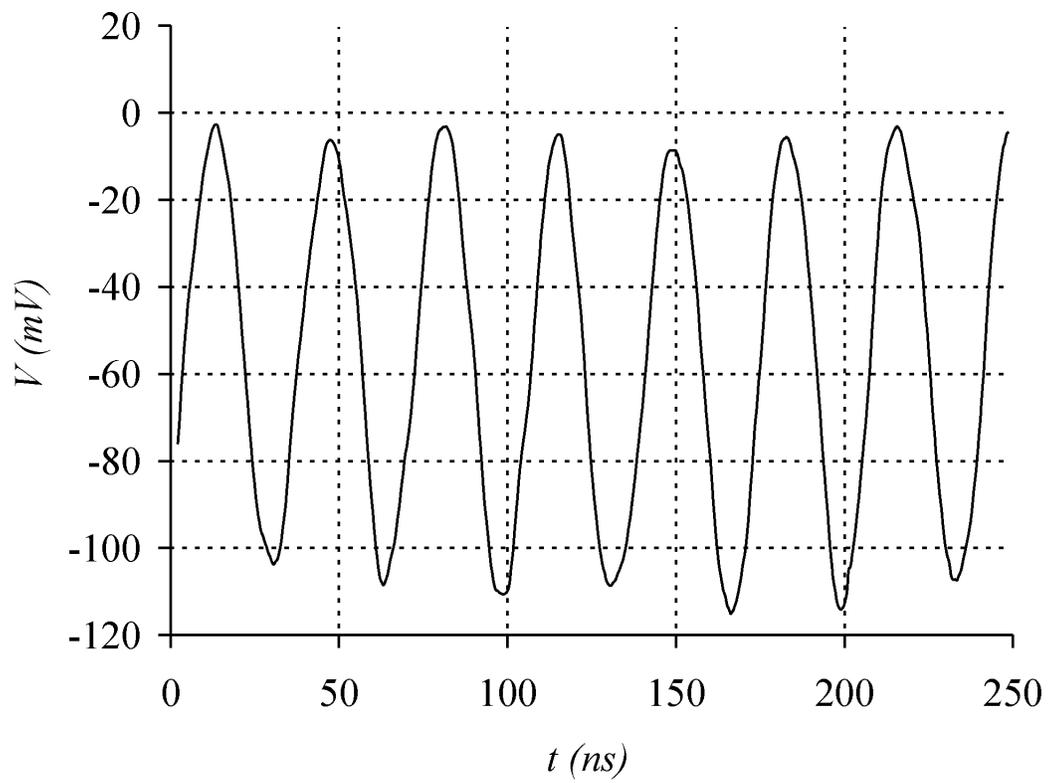

BernalFig3b

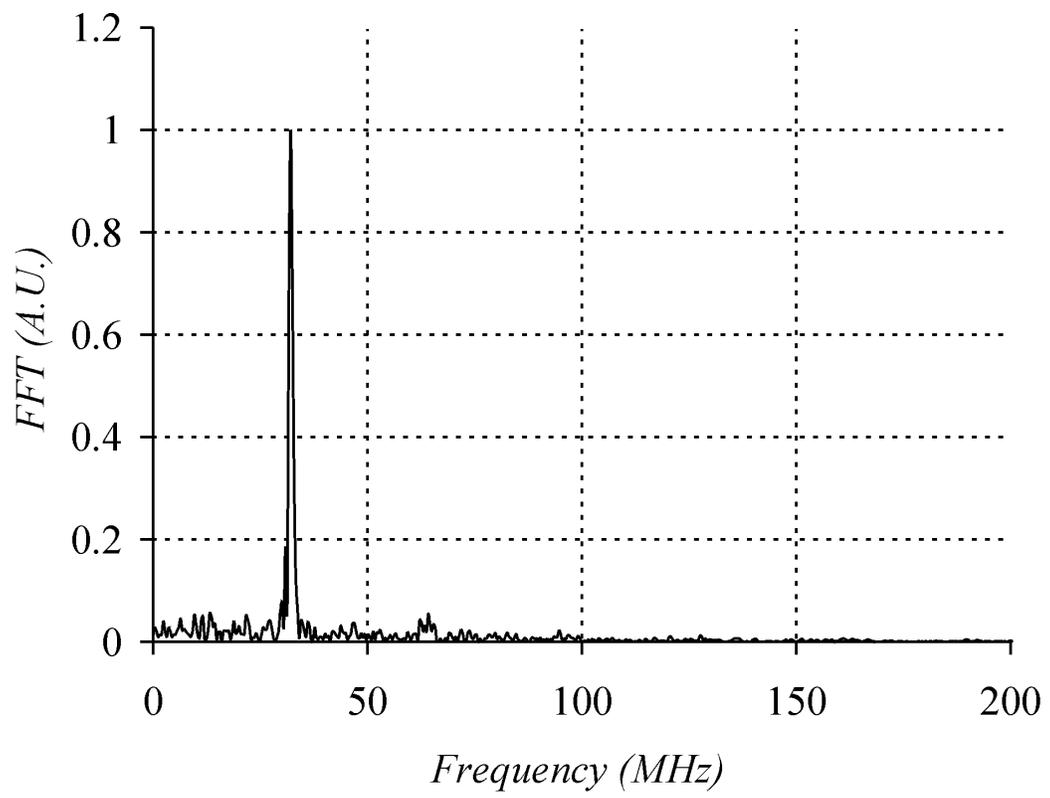

BernalFig4a

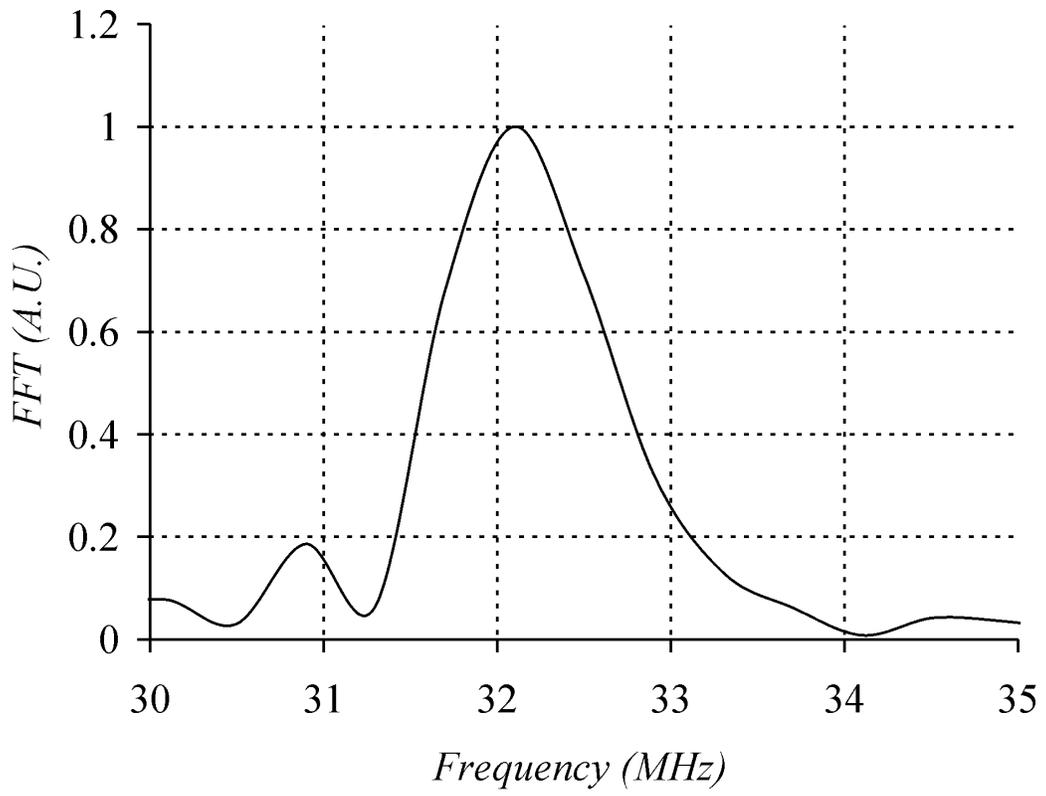

BernalFig4b

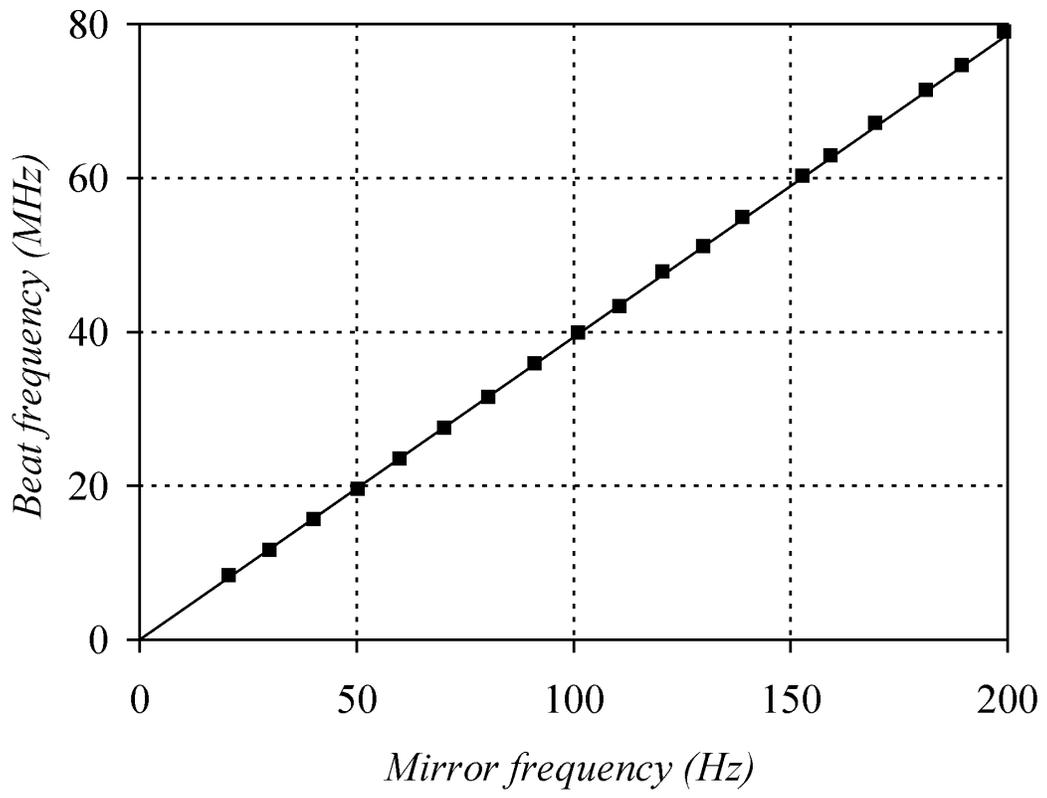

BernalFig5